\documentclass[twocolumn,aps,pra,preprintnumbers,10pt]{revtex4-1}
\pdfoutput=1

\usepackage{graphicx}
\usepackage{dcolumn}

\usepackage{epstopdf}
\usepackage{amsmath}
\usepackage{amssymb}
\usepackage{amsfonts}
\usepackage{physics}
\usepackage{siunitx}
\usepackage{ulem}
\usepackage[colorlinks,linkcolor=blue,citecolor=blue,urlcolor=blue]{hyperref}
\usepackage[usenames,dvipsnames]{color}
\begin{document}

\title{Measurements of spin properties of atomic systems in and out of equilibrium via noise spectroscopy}

\author{Maheswar Swar}
\email{mswar@rri.res.in}
\author{Dibyendu Roy}
\author{Dhanalakshmi D}
\author{Saptarishi Chaudhuri}
\email{srishic@rri.res.in}
\author{ Sanjukta Roy}
\email{sanjukta@rri.res.in}
\author{Hema Ramachandran}
\affiliation{Raman Research Institute, C. V.  Raman Avenue, Sadashivanagar, Bangalore-560080, India.}
\date{\today}

\begin{abstract}
 We explore the applications of spin noise spectroscopy (SNS) for detection of the spin properties of atomic ensembles in and out of equilibrium. In SNS, a linearly polarized far-detuned probe beam on passing through an ensemble of atomic spins acquires the information of the spin correlations of the system which is extracted using its time-resolved Faraday-rotation noise. We measure various atomic, magnetic and sub-atomic properties as well as perform precision magnetometry using SNS in rubidium atomic vapor in thermal equilibrium.  Thereafter, we manipulate the relative spin populations between different ground state hyperfine levels of rubidium by controlled optical pumping which drives the system out of equilibrium. We then apply SNS to probe such spin imbalance non-perturbatively. We further use this driven atomic vapor to demonstrate that SNS can have better resolution than typical absorption spectroscopy in detecting spectral lines in the presence of various spectral broadening mechanisms.
\end{abstract}

\maketitle
\section{Introduction}
\label{intro}

Control of spin population and its simultaneous non-destructive detection play a crucial role in diverse scientific fields such as atom interferometry \cite{Cronin2009}, precision magnetometry \cite{Brask2015}, atomic clocks \cite{Wynands2005}, quantum simulation \cite{Georgescu2014} and quantum information processing \cite{Monroe2002}. While external magnetic fields and optical pumping can be used to manipulate the spin polarization and population in an atomic system, spin noise spectroscopy (SNS) \cite{Muller2010a, Zapasskii2013, Sinitsyn2016} provides a means of the detection of such spin coherences by probing spontaneous spin fluctuations of the system via an off-resonant laser beam, and is relatively non-perturbative as compared to the traditional absorption spectroscopy. In this work, we show that SNS can measure different spin properties of rubidium (Rb) vapor not only in equilibrium systems, but also in systems driven out of equilibrium. 

The random fluctuations over space and time are prevalent in a wide variety of physical systems, and they can be a valuable resource for probing the characteristic nature and internal structure of the systems. Examples of such fluctuations or noise include the Brownian motion of pollen grains in water, the Johnson noise due to thermal agitation of the electrons in an electrical conductor \cite{Johnson1927}, the intensity fluctuations in the emission of random lasers \cite{Sharma06,Gomes2016} and the stochastic fluctuations in clonal cellular constituents \cite{Munsky2009, Swaminathan1982}.  

The optical SNS technique has been developed by Aleksandrov and Zapasskii \cite{Aleksandrov1981} to passively probe intrinsic spin fluctuations or magnetization noise in a thermal ensemble of spins. These spin ensembles can be made of electron spins in atomic systems and spins of electrons or holes in semiconductors and other solid-state materials. A study of SNS in alkali atomic vapor of Rb and potassium (K) in thermodynamic equilibrium was carried out by \cite{Crooker2004}, which indicated that the electronic and nuclear g-factors, isotope abundance ratios, nuclear moments and hyperfine splittings could be measured. The first successful application of SNS to a solid-state system was performed by \cite{Oestreich2005}, for measuring the electron's Lande g-factor and spin relaxation time in a $n$-doped GaAs semiconductor. There has been significant progress in the recent years to extend the applicability of SNS \cite{Mihaila2006, Katsoprinakis2007, Muller2008, Crooker2009, Crooker2010, Muller2010b, Yan2012, Zapasskii2013b, Pershin2013, Berski2013, Dahbashi2014, Poltavtsev2014, Yang2014, Roy2015, Lucivero2016, Sterin2018}.

First, we perform the SNS of Rb atoms in thermal equilibrium. We demonstrate accurate measurements of several physical quantities such as electron's g-factor, nuclear g-factor, isotope abundance ratios, and develop precision magnetometry with our thermal Rb vapor in the presence of a static magnetic field perpendicular to probe laser. While prior measurements have been reported \cite{Crooker2004, Mihaila2006} of several of these quantities using SNS technique, we are able to refine some of these estimates especially for isotope abundance ratios and nuclear g-factor. We then apply an optical pumping beam to control relative spin population in the ground state hyperfine levels of Rb atoms. This is realized by an on-resonance pump beam nearly co-propagating with the far-detuned probe beam. The optical pumping drives the system out of equilibrium. We then show that SNS can be used to determine spin imbalance in different ground state hyperfine levels without disturbing the non-equilibrium steady-state of the system. We also show that the spin noise (SN) spectra from the optically pumped atoms have better resolution than typical absorption spectra from the same system. Therefore, the SNS can be used in resolving spectral lines of a non-equilibrium system in the presence of various spectral broadening mechanisms.       

This paper is organized as follows: We provide a brief theoretical description of SNS in Sec.~\ref{theory}. In Sec.~\ref{setup}, the experimental set-up and measurement methods are presented in detail. In Sec.~\ref{equi}, we describe measurements and results of SNS in Rb vapor in thermal equilibrium.  The results of SNS for optically pumped Rb vapor are given in Sec.~\ref{outEq}. The final Sec.~\ref{conc} comprises a conclusion and an outlook.

\section{Theory}
\label{theory}

Let us consider an ensemble of non-interacting spins in thermal equilibrium at some temperature. The equilibrium is achieved through interactions between these spins and a thermal bath surrounding them. The presence of the thermal bath induces fluctuations in spin polarization over time. Nevertheless, the time-averaged value of the spin polarization or magnetization, $\langle M(t)\rangle_{T\rightarrow \infty}$ ($T$ is the total averaging time) along any arbitrary quantization axis is zero for a paramagnetic system. However, the variance of the magnetization is non-zero. Within the optical SNS technique, a linearly polarized laser light on passing through such a paramagnetic sample can passively detect these magnetization fluctuations along the light propagation in its time-resolved Faraday rotation noise \cite{Crooker2004, Zapasskii2013}. Such detection is feasible as the magnetization fluctuations in a paramagnetic sample alter its optical properties which lead to Faraday rotation noise. The probe beam is kept far-detuned (with a detuning $\delta_p$) from any allowed optical transition of the medium to ensure negligible scattering by the medium making SNS a relatively non-invasive technique.

The intrinsic fluctuations of the spin polarization reveal the characteristic relaxation times of the system. In the presence of a constant magnetic field, the spontaneous spin polarization precesses at the Larmor frequency about the magnetic field. Assuming a single exponential spin relaxation time $T_2$ and a magnetic field $B_{\bot}$ being orthogonal to the probe laser propagation ($\widehat{z}$), we can obtain the temporal correlation of magnetization along the probe beam from the Bloch equation:
\begin{equation}
\langle M_z(t)M_z(0)\rangle \propto \cos(\nu_{L} t)e^{-t/T_2},\label{Eqn:Lorentz_SNS}
\end{equation}
where the Larmor frequency $\nu_{L}=g_F\mu_BB_{\bot}/h$, $g_F$ is the $g$-factor of the hyperfine $F$-levels, $\mu_B$ is the Bohr magneton and $h$ is the Planck's constant.   

The measured Faraday rotation fluctuation $\langle \theta_F(t)\theta_F(0)\rangle$ ($\theta_F(t)$ is the Faraday rotation angle at time $t$) is a direct probe of the magnetization fluctuation $\langle M_z(t)M_z(0)\rangle$ of the system in thermal equilibrium and its Fourier transform to spectral frequency $\nu$ is the power spectral density $P(\nu)$ of the spin noise. Therefore,
\begin{eqnarray}
P(\nu>0)&=&\int_{0}^{\infty}dt\: \cos(\nu t)\langle \theta_F(t)\theta_F(0)\rangle \nonumber \\ &\propto& \frac{1/T_2}{(\nu-\nu_L)^2+1/T_2^2}, \label{SNpower}
\end{eqnarray}
where we have used Eq.~\ref{Eqn:Lorentz_SNS} in the last line. So, the SN power spectrum has a Lorentzian lineshape centred at $\nu_L$ in frequency domain (refer to the peaks in the SN amplitude spectrum $\sqrt{P(\nu>0)}$ in Fig.~\ref{fig:Experimental_set-up}(b) for $^{87}$Rb or $^{85}$Rb) and its full width at half maxima (FWHM) is proportional to $1/T_2$.

The energy $E_{F,m_F}$ of hyperfine $F$-levels for alkali atoms in an arbitrary magnetic field $B_{\bot}$ has an exact expression following Breit and Rabi \cite{Mockler1961},
\begin{eqnarray}
\label{Eqn:Breit-Rabi}
E_{F=I\pm\frac{1}{2},m_F}&=&-\frac{h\Delta_{\text{hf}}}{2(2I+1)} + g_I\mu_B B_{\bot} m_F \nonumber \\
&\pm& \frac{h\Delta_{\text{hf}}}{2} \sqrt{1+\frac{4m_F}{2I+1}x+x^2},
\end{eqnarray}
where $h\Delta_{\text{hf}}$, $g_I$ and $m_F$ are the zero-field hyperfine separation between the levels $F=I+\frac{1}{2}$ and $F=I-\frac{1}{2}$, the nuclear $g$-factor and the magnetic quantum number, respectively. Here, $x=(g_J-g_I)\mu_B B_{\bot}/(h\Delta_{\text{hf}})$ where $g_J$ is the Lande $g$-factor and the nuclear spin $I = 3/2$ for $^{87}$Rb. Since, the SNS detects the spin coherences between different Zeeman sub-levels ($\bigtriangleup m_F=\pm 1$), the frequencies of different magnetic resonance peaks have a nonlinear dependence on $B_{\bot}$ \cite{steck87}.

 The integrated SN power over frequency, $\chi \equiv \int d\nu P(\nu >0 )$, depends on the probe detuning as $\chi \propto \delta_p^{-2}$ \cite{Mihaila2006} and is symmetric about the atomic resonance frequency for a far-detuned probe beam (where $\delta_p \gg \Gamma$, $\Gamma$ being the width of the absorption spectra). However, this integrated SN power $\chi$ becomes asymmetric over $\delta_p$ for $\delta_p \approx 0$ due to the non-vanishing coherences between the ground and excited state hyperfine levels of the atoms \cite{Horn2011}. This asymmetry in $\chi$ is shown later in this paper in Fig.~\ref{fig:SNA_detuning}(b,d). The asymmetry in $\chi$ also depends on the homogeneous and inhomogeneous broadening present in the medium \cite{Horn2011, Zapasskii2013b}.

\section{Experimental set-up and methods}
\label{setup}

\begin{figure}
\centering
\includegraphics[scale=0.4]{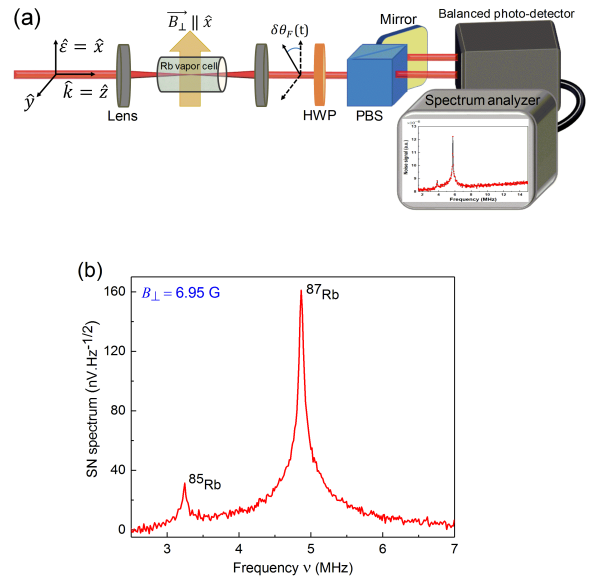}
\caption{(a) Schematic of the experimental set-up for measuring spin noise (SN) spectrum. A probe beam along $\hat{z}$ is focused by a plano-convex lens before entering the vapor cell. The transmitted probe beam is sent through a polarimetric set-up comprising of a half-wave plate (HWP) and a polarizing beam splitter (PBS), and then it is collected by a balanced photo-detector which is connected to a spectrum analyzer. A constant magnetic field $B_\bot$ along $\hat{x}$ is applied on the atomic vapor using two magnetic coils in Helmholtz configuration. (b) A typical SN amplitude spectrum at $B_\bot = 6.95$ G and vapor cell temperature of $105^{\circ}$C is shown for a probe detuning $\delta_p=-10.2$ GHz. The stronger and weaker peaks are identified with $^{87}$Rb $(|g_F| = 1/2)$ and $^{85}$Rb $(|g_F| = 1/3)$ respectively.}

\label{fig:Experimental_set-up}
\end{figure}

\begin{figure}
\centering
\includegraphics[scale=0.38]{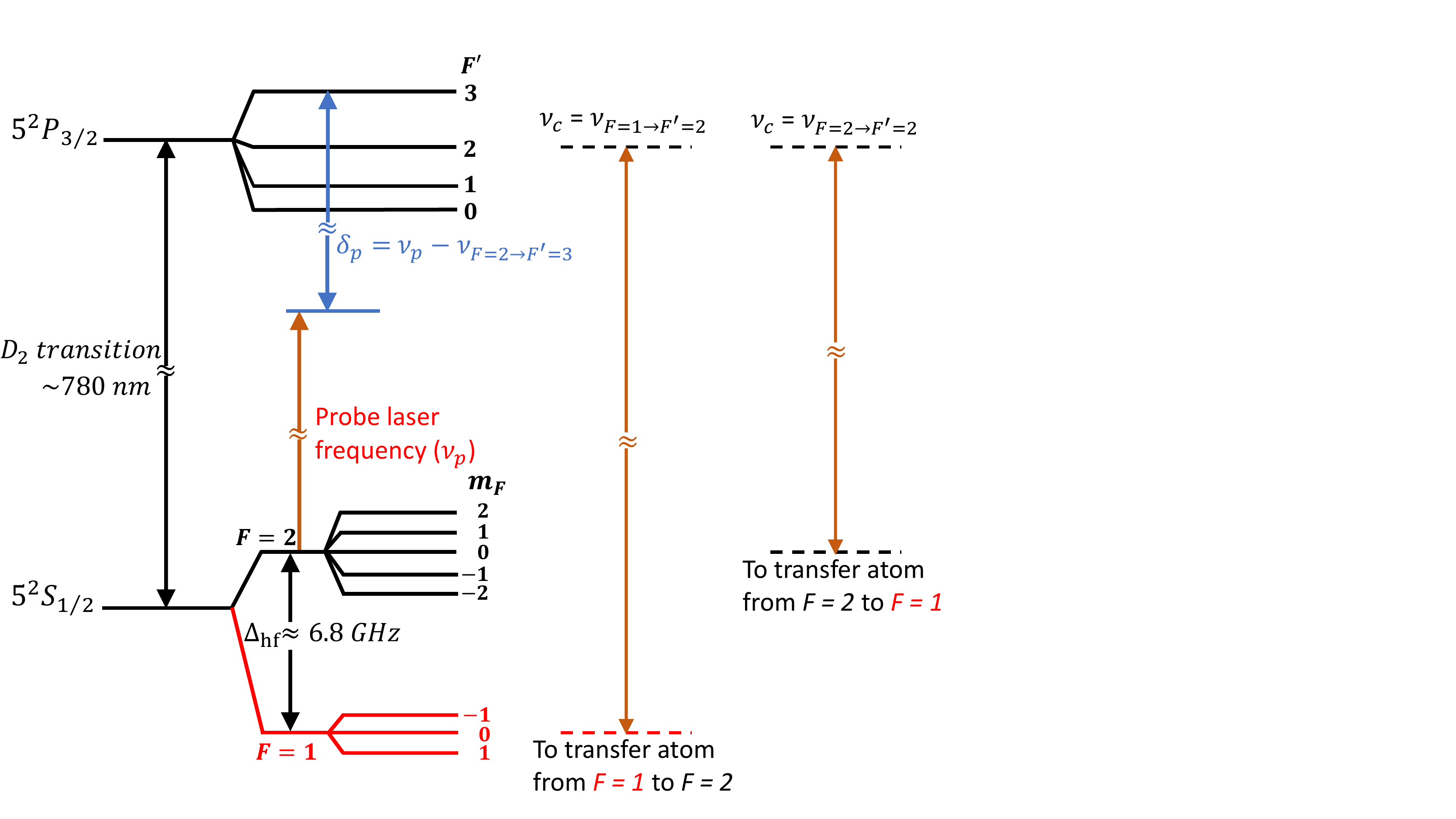}
\caption{Energy level diagram for $D_2$ transition of $^{87}$Rb atoms. The two ground state hyperfine levels ($F = 1$ and $F = 2$) are separated by $\sim 6.8$ GHz. The probe laser frequency $\nu_p$ is detuned by $\delta_p$ from the $F = 2 \rightarrow F' = 3$ transition, i.e., $\delta_p = \nu_{p} - \nu_{F = 2 \rightarrow F' = 3}$, where $\nu_{F = 2 \rightarrow F' = 3}$ is the frequency between $F = 2$ and $F' = 3$ hyperfine levels. The Zeeman sub-levels of the ground state hyperfine levels  in the presence of external magnetic field are depicted. The selected transitions for optical pumping of atoms by a control laser of frequency $\nu_c$ are also shown.}
\label{fig:Energy_Diagram}
\end{figure}

The schematic of the experimental set-up is shown in Fig.~\ref{fig:Experimental_set-up}(a). A linearly polarized probe laser beam with tunable frequency $\nu_p$ is sent through a 20 mm long glass cell containing enriched $^{87}$Rb vapor. This probe beam is derived from a grating stabilized external cavity diode laser in Littrow configuration having an instantaneous linewidth below 1 MHz. The probe beam with a Gaussian profile is focused inside the atomic vapor to a $1/e^2$ waist size of 45 $\mu$m at the focal plane and a Rayleigh range of 4 mm. In order to study the dependence of the SN spectra on the probe beam detuning $\delta_p$, the probe frequency $\nu_p$ is varied over a large range of frequencies ($\sim$ 25 GHz). The relevant energy-levels of $^{87}$Rb are depicted in Fig.~\ref{fig:Energy_Diagram}. Special care is taken so that the laser operates in a mode-hop free regime. The frequency $\nu_p$ of the probe beam is measured using a commercial wavelength meter (HighFinesse, model-WSU2) with a relative accuracy of $\pm$ 1 MHz.

The glass cell is filled with neon buffer gas at a pressure of $200$ mbar in order to make the medium diffusive for Rb atoms. This increases the transverse transit time of the atoms across the probe beam from $\sim$ 200 ns to $\sim$ 100 $\mu$s providing sufficient time for acquiring time resolved Faraday rotation signal for the accurate detection of the atomic properties.  We collect each real-time Faraday rotation signal for relatively longer time duration than the transit time. However, the spin life-time of Rb atoms at room temperature is of the order of milliseconds, \cite{Arditi1964} which is much longer than the above transverse transit time ($\sim$ 100 $\mu$s). The inert gas neon is chosen because the collisions between Rb and neon do not change the spin state of the Rb atoms. The vapor cell is connected to a controllable heater in order to vary the number density of the atomic spin ensemble.

The atoms are subjected to a uniform, constant magnetic field ($B_\bot \hat{x}$) perpendicular to the direction of propagation of the probe beam ($\hat{z}$). This field, generated using two circular coils in Helmholtz configuration, is uniform along the length of the cell within $\pm 0.4\%$. The entire experimental set-up is shielded with a mild-steel box ($\mu/\mu_0 = 2000$) to avoid any unwanted stray magnetic field. 
 
The probe beam after passing through the glass cell is separated into $s$- and $p$-polarized components using a polarization sensitive set-up as shown in Fig.~\ref{fig:Experimental_set-up}(a). The two components are then fed into the two ports of a balanced photo detector (Newport model no. 1807-FS) that has a 3 dB bandwidth of 80 MHz and a good common mode rejection ratio of $25$ dB. The output of the balanced detector is directly connected to a spectrum analyzer (Agilent CSA Spectrum Analyzer Model no. N1996A, frequency range $100$ kHz - $3$ GHz) whose resolution bandwidth is adjusted between $100$ Hz to $1$ kHz. The spectrum analyzer was set on continuous averaging mode for two to five minutes for recording various SN spectra presented in this paper.

\section{Measurements and Results in equilibrium}
\label{equi}

A typical spin noise spectrum of Rb atoms in thermal equilibrium at low $B_\bot$ ($=6.95$ G) is presented in Fig.~\ref{fig:Experimental_set-up}(b). This signal was obtained with the vapor cell at 105$^{\circ}$C and a $p$-polarized ($\hat{\varepsilon}\parallel\hat{x}$), 300 $\mu W$ probe beam. The probe beam is red-detuned by 10.2 GHz with respect to the $D_2$ transition (at $\sim$ 780 nm) of $^{87}$Rb, $F=2\rightarrow F'=3$. Two distinct noise peaks, one at 3.24 MHz and another at 4.87 MHz, are observed and identified as arising due to spin fluctuations among the intra-hyperfine Zeeman sublevels ($\bigtriangleup F=0$, $\bigtriangleup m_F=\pm1$) of  $^{85}$Rb and $^{87}$Rb, respectively. The photon shot noise background of $\sim$ 350 nV.Hz$^{-1/2}$ is subtracted from the noise spectrum. The observed SN peaks are very narrow ($<$ 100 kHz) and the peak positions (which occur at the $\nu_L$) can be detected with a precision of $\sim 1$ part in $10^5$. This makes it possible to employ SNS for a variety of precision measurements as we demonstrate  in the following sections.

\subsection{Measurements of g-factors and isotope abundance}

\begin{figure}
\centering
\includegraphics[scale=0.42]{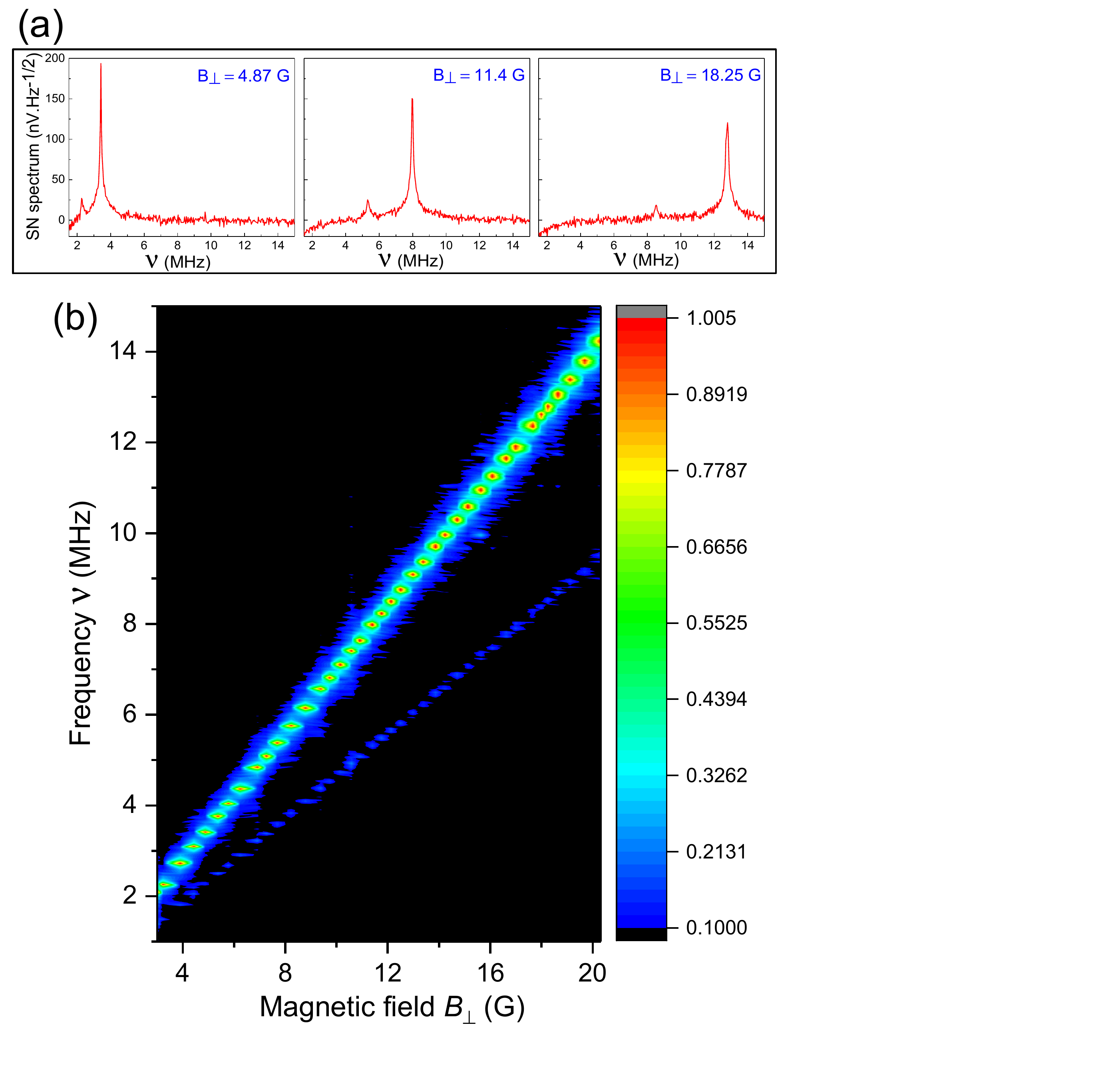}
\caption{ (a) Spin noise (SN) spectra at various $B_\bot$ are presented. The linear shift of the noise peaks with $B_\bot$ suggests a linear Zeeman effect of the ground  state hyperfine levels in that $B_\bot$ range. (b) A 2-D false color mapping shows the SN peak positions for $^{85}$Rb and $^{87}$Rb as a function of $B_\bot$. The bright (faint) trace is for $^{87}$Rb ($^{85}$Rb) SN signal. The noise signal strength of $^{87}$Rb and $^{85}$Rb for each spectrum is plotted after normalizing the signal by the SN peak strength of $^{87}$Rb. The slopes of these traces reveal  $|g_F|$ of Rb isotopes.}
\label{fig:SNS Signal}
\end{figure}

Fig.~\ref{fig:SNS Signal}(a) shows SN spectra at three representative values of $B_\bot$ illustrating that the two noise peaks, corresponding to $^{87}$Rb and $^{85}$Rb, shift in positions with $B_\bot$. Fig.~\ref{fig:SNS Signal}(b) gives the variation in the position of these noise peaks as a function of $B_\bot$. The bright (faint) trace corresponds to the spin noise peak positions of $^{87}$Rb ($^{85}$Rb) atoms. The linear dependence of the peak positions  on $B_\bot$ indicates that the system is in the linear Zeeman regime. The slope of these lines  give a measure of the g-factor $|g_F|$ for the ground state hyperfine levels. The g-factors  obtained from our measurements are $|g_F| = 0.500(\pm 0.001)$ for $^{87}$Rb and $|g_F| = 0.333(\pm 0.001)$ for $^{85}$Rb which are in excellent agreement with the theoretical values.

Our measurements were made with an enriched $^{87}$Rb vapor cell. Traditional absorption spectroscopy did not show the presence of the isotope $^{85}$Rb. However, SN spectra clearly indicate the presence of both isotopes in the cell. From the ratio of the total integrated SN power ($\chi$) of the two peaks, we estimate an abundance ratio of $^{87}$Rb $:$ $^{85}$Rb $= 11:1$ from Fig.~\ref{fig:Experimental_set-up}(b). This shows that SNS is a very sensitive technique for detecting abundance ratios of various isotopes with high precision even when present in minute quantities.

\subsection{Precision magnetometry}

\begin{figure}
\centering
\includegraphics[scale=0.42]{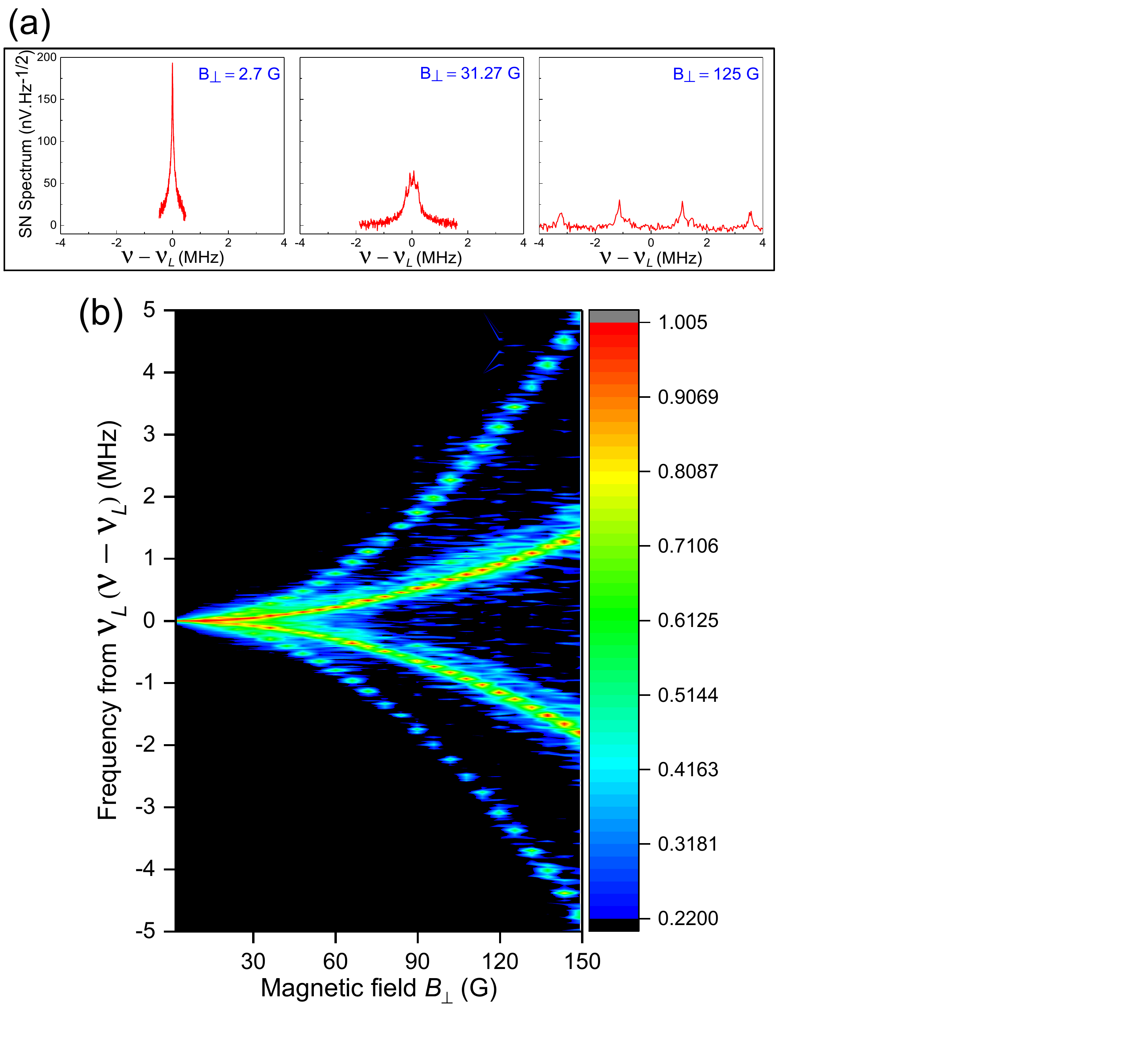}

\caption{ Broadening and splitting of the spin noise (SN) spectrum with increasing $B_\bot$. (a) SN spectra for $^{87}$Rb at $B_\bot = 2.7$ G, $31.27$ G, $125$ G. The  origin of the frequency in these spectra is shifted to the central Larmor frequency $\nu_L=g_F\mu_BB_{\bot}/h$.  The other parameters are: probe power = $400 \mu$W, $\delta_p = -10.6$ GHz, cell temperature = $105^{\circ}$C. (b) Visual realization of the nonlinear Zeeman effect of ground state hyperfine levels with increasing $B_\bot$. Each spectrum is normalized by the strongest peak in the SN signal.} 
\label{fig:Quadratic_Zeeman}
\end{figure}

On increasing the magnitude of the applied magnetic field ($B_\bot$), the spin noise spectra is observed (Fig.~\ref{fig:Quadratic_Zeeman}(a)) to broaden ($B_\bot\sim 25-40$ G) and to split into well-resolved peaks at even higher $B_\bot$ ($> 60$ G). At such high fields, the system is clearly in non-linear Zeeman regime (Eq.~\ref{Eqn:Breit-Rabi}). A false-color mapping of the measured nonlinear Zeeman splitting of the ground state hyperfine levels of $^{87}$Rb atoms as a function of $B_\bot$ is shown in Fig.~\ref{fig:Quadratic_Zeeman}(b). 

\begin{figure}
\centering
\includegraphics[scale=0.4]{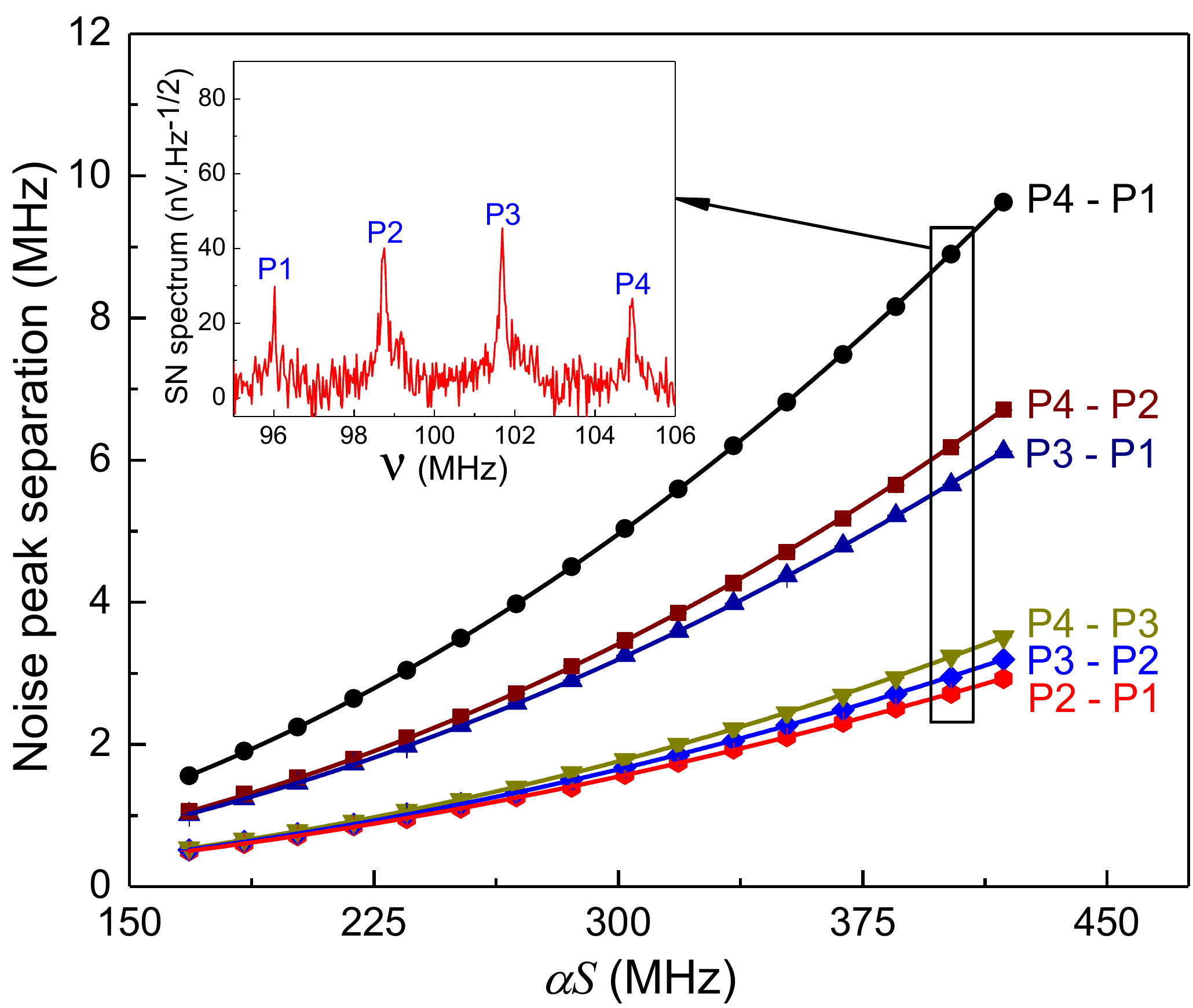}
\caption{Measurement of magnetic field $B_{\bot}$ and $\Delta_{\text{hf}}$ using spin noise spectroscopy. $P1,P2,P3$ and $P4$ indicate the position of spin noise peaks (corresponding to Zeeman sub-levels of $F = 2$) as can be seen from the raw data in the inset. The measured frequency separation between different noise peaks are plotted against measured $\alpha S$ (refer to the text). The bold lines are obtained from the Breit-Rabi formula in Eq.~\ref{Eqn:Breit-Rabi1} with known parameters for error analysis in magnetic field measurements. The experimental parameters are the same as in Fig.~\ref{fig:Quadratic_Zeeman}.
}
\label{fig:magnetometry}
\end{figure}

The individual noise peaks in the SN spectrum for a higher $B_\bot$ $(> 60$ G$)$ can be identified as the transitions between different Zeeman sub-levels of the ground state hyperfine levels. These are shown in the inset of Fig.~\ref{fig:magnetometry} where $P1$ denotes the magnetic resonance frequency between ($F = 2, m_F = 2) \leftrightarrow (F = 2, m_F = 1$) and $P2$ for the magnetic resonance frequency between ($F = 2, m_F = 1) \leftrightarrow (F = 2, m_F = 0$) and so on. Fig.~\ref{fig:Quadratic_Zeeman}(b) shows nonlinear dependence of each noise peak frequency on $B_\bot$. However, the sum of all four noise peak frequencies, 
\begin{eqnarray}
\label{Eqn:S}
S=P1 + P2 + P3 + P4 = \frac{\mu_B}{h}(3 g_I + g_J) B_{\bot}, 
\end{eqnarray}
depends linearly on $B_{\bot}$. Using the values of $P1, P2, P3, P4$ determined from the measured SN spectrum one can estimate $B_{\bot}$ using Eq.~\ref{Eqn:S}, substituting the values of $\mu_B,h,g_I,g_J$, which are already known to high precision. As the observed SN peaks are extremely narrow (FWHM $< 100$ kHz) and the peak positions can be determined with an accuracy of one part in $10^5$, we can therefore measure an external magnetic field within that same order of relative error, of one part in $10^5$, for the range of $B_{\bot}$ where noise peaks are separable. Thus SNS provides a simple means of precision magnetometry.

Now we rewrite Eq.~\ref{Eqn:Breit-Rabi} for $^{87}$Rb as
\begin{eqnarray}
\label{Eqn:Breit-Rabi1}
E_{F=2,m_F}&=&-\frac{h\Delta_{\text{hf}}}{8} + \frac{h g_I}{(g_J-g_I)} (\alpha S) m_F\nonumber \\ 
&+&\frac{h}{2} \sqrt{\Delta_{\text{hf}}^2+\Delta_{\text{hf}}(\alpha S)m_F+(\alpha S)^2},
\end{eqnarray}
where $\alpha=(g_{J}-g_{I})/(g_{J}+3g_{I})$. In Fig.~\ref{fig:magnetometry}, we plot, as a function of $\alpha S$, the noise peak separations (($P4-P1$), ($P4-P2$) ... ($P2-P1$)) calculated from Eq.~\ref{Eqn:Breit-Rabi1} using known values of $h,g_I,g_J,\Delta_{\text{hf}}$. Superposed on the plot are the experimentally obtained noise peak separations shown as solid symbols. Then, we note down the x-errors between the experimentally obtained peak separations and those from the calculated curves. The root-mean-square value of the x-errors gives an estimate of the error in measuring the external magnetic field. We find that the error is within $500~\mu$G in our measurement range of $B_{\bot}$ between $60$ G to $150$ G. This accuracy surpasses the standard Hall probe based magnetometers by nearly two orders of magnitude. Moreover, this high precision measurement of magnetic field is an in-situ detection, without requiring the physical placement of a separate probe.

In Fig.~\ref{fig:magnetometry}, we have also fitted the experimentally obtained values of ($P4-P1$), ($P4-P2$) ... ($P2-P1$) using Eq.~\ref{Eqn:Breit-Rabi1} keeping $\Delta_{\text{hf}}$ as a free parameter. From these fittings, we extract the value of the zero-field hyperfine constant $\Delta_{\text{hf}} \sim 6805.5 (\pm 7.2)$ MHz.

\subsection{Nuclear g-factor from SNS}

\begin{figure}
\centering
\includegraphics[scale=0.4]{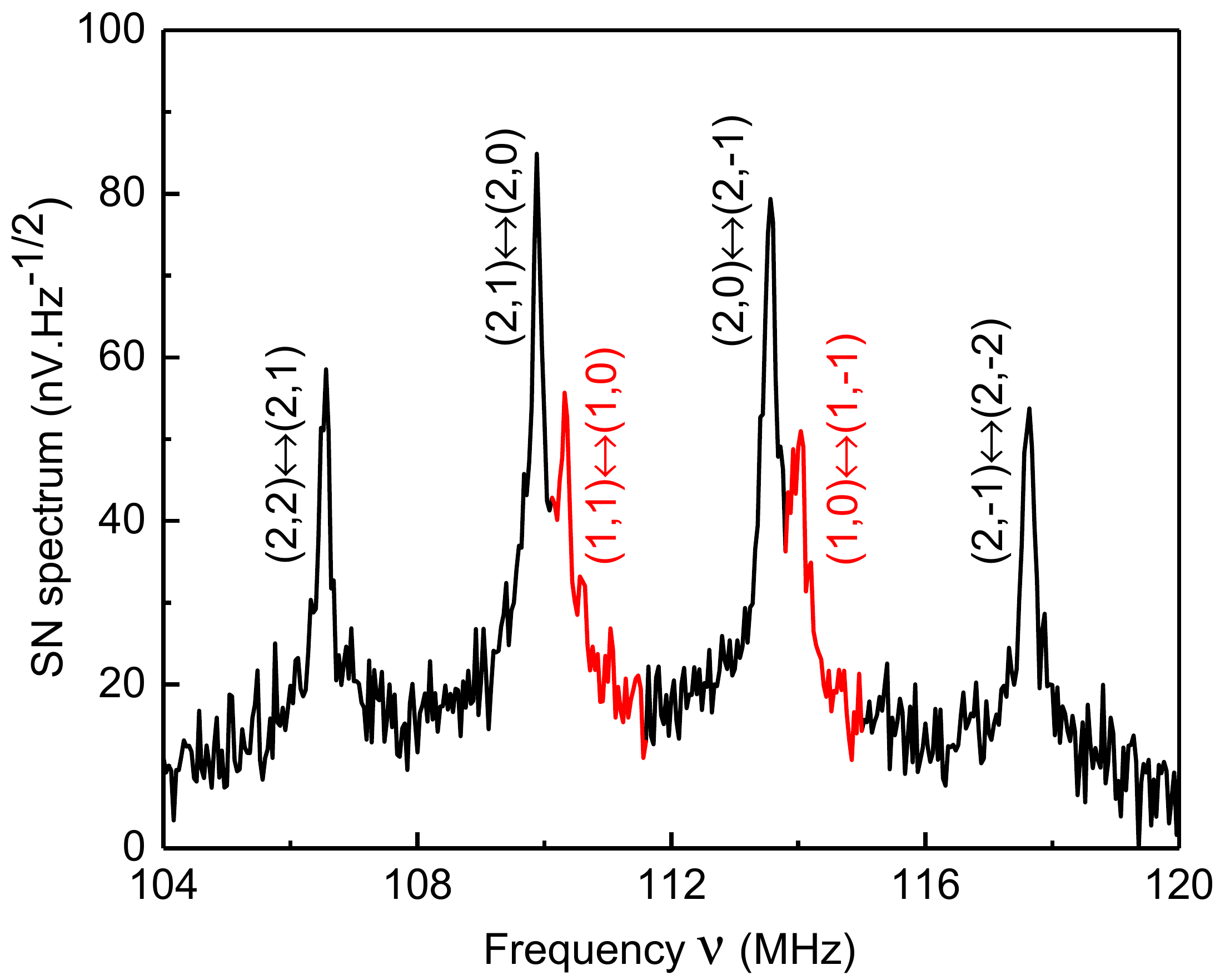}
\caption{Spin noise spectrum with resolved all-allowed Zeeman transitions ($\bigtriangleup F = 0$, $\bigtriangleup m_F = \pm1$) of ground state hyperfine levels in $^{87}$Rb. The parameters are $B_\bot = 160$ G, probe power = $750~\mu$W, $\delta_p = -10.6$ GHz, cell temperature = $105^{\circ}$C. The value of nuclear g-factor $(g_I)$ is precisely obtained and reported in the text by measuring the separation between $(2,1) \leftrightarrow (2,0)$ and $(1,1) \leftrightarrow (1,0)$ (also  $(2,0) \leftrightarrow (2,-1)$ and $(1,0) \leftrightarrow (1,-1)$) in a series of measurements.}
\label{fig:NuclearZeeman}
\end{figure}

In the presence of $B_{\bot}$, the energy separations between similar magnetic (Zeeman) transitions from different hyperfine ground states $(F, m_F)$ such as,  $(2,1) \leftrightarrow (2,0)$ and $(1,1) \leftrightarrow (1,0)$ or $(2,0) \leftrightarrow (2,-1)$ and $(1,0) \leftrightarrow (1,-1)$ in Fig.~\ref{fig:Energy_Diagram}, are determined by the second term in Eq.~\ref{Eqn:Breit-Rabi} arising out of the nuclear spin. 
However, since the value of nuclear g-factor ($g_I$) is small, the contribution of this term is negligible for low magnetic fields. Therefore, the SN peaks from $(2,1) \leftrightarrow (2,0)$ and $(1,1) \leftrightarrow (1,0)$ are almost unresolved for $B_{\bot}< 150$ G in our case. At high magnetic fields ($>150$ G), we can resolve the SN peaks from all available Zeeman transitions when their separations are more than the width of the individual peaks. Six distinct SN peaks from the allowed $\bigtriangleup F = 0$, $\bigtriangleup m_F = \pm1$ transitions of $^{87}$Rb are observed in Fig.~\ref{fig:NuclearZeeman} at $B_{\bot} = 160$ G. The value of $g_I$ can be precisely obtained by measuring the separation between $(2,1) \leftrightarrow (2,0)$ and $(1,1) \leftrightarrow (1,0)$ (also  $(2,0) \leftrightarrow (2,-1)$ and $(1,0) \leftrightarrow (1,-1)$). From a series of such measurements, the experimentally estimated $g_I$ for $^{87}$Rb in our experiment is $-0.00100627(\pm 0.00002558)$, where the quantity in the bracket refers to the $1\sigma$ error. 

\section{Measurements and Results in optically pumped vapor}
\label{outEq}

Thus far, we have explored SNS in an equilibrium thermal vapor. We now report the measurements on out-of-equilibrium systems where we use optical fields to manipulate spin populations in the different ground state hyperfine levels. Recently, the SNS was employed to detect couplings and correlations between different spin coherences in a non-equilibrium atomic vapor \cite{Glasenapp2014}. In the experiment in \cite{Glasenapp2014}, the Zeeman sub-levels of the ground state hyperfine levels of $^{41}$K are driven by a weak radio frequency magnetic field which brings the vapor out of equilibrium. In contrast, we apply an optical control beam between the ground state hyperfine levels and the excited state hyperfine levels to drive as well as control the spin populations. The control beam is linearly polarized and almost co-propagating with the probe beam.

In our experiment, the control beam is derived from an independent tunable external cavity diode laser. The Rb atoms in the vapor cell are optically pumped to the desired ground state hyperfine levels by tuning the frequency $\nu_c$ and the intensity $I_c$ of the pump laser. Substantial modifications of the SNS signals are observed depending upon the relative ground state hyperfine level populations of the vapor.

\subsection{Detection of spin imbalance}

\begin{figure}
\centering
\includegraphics[scale=0.43]{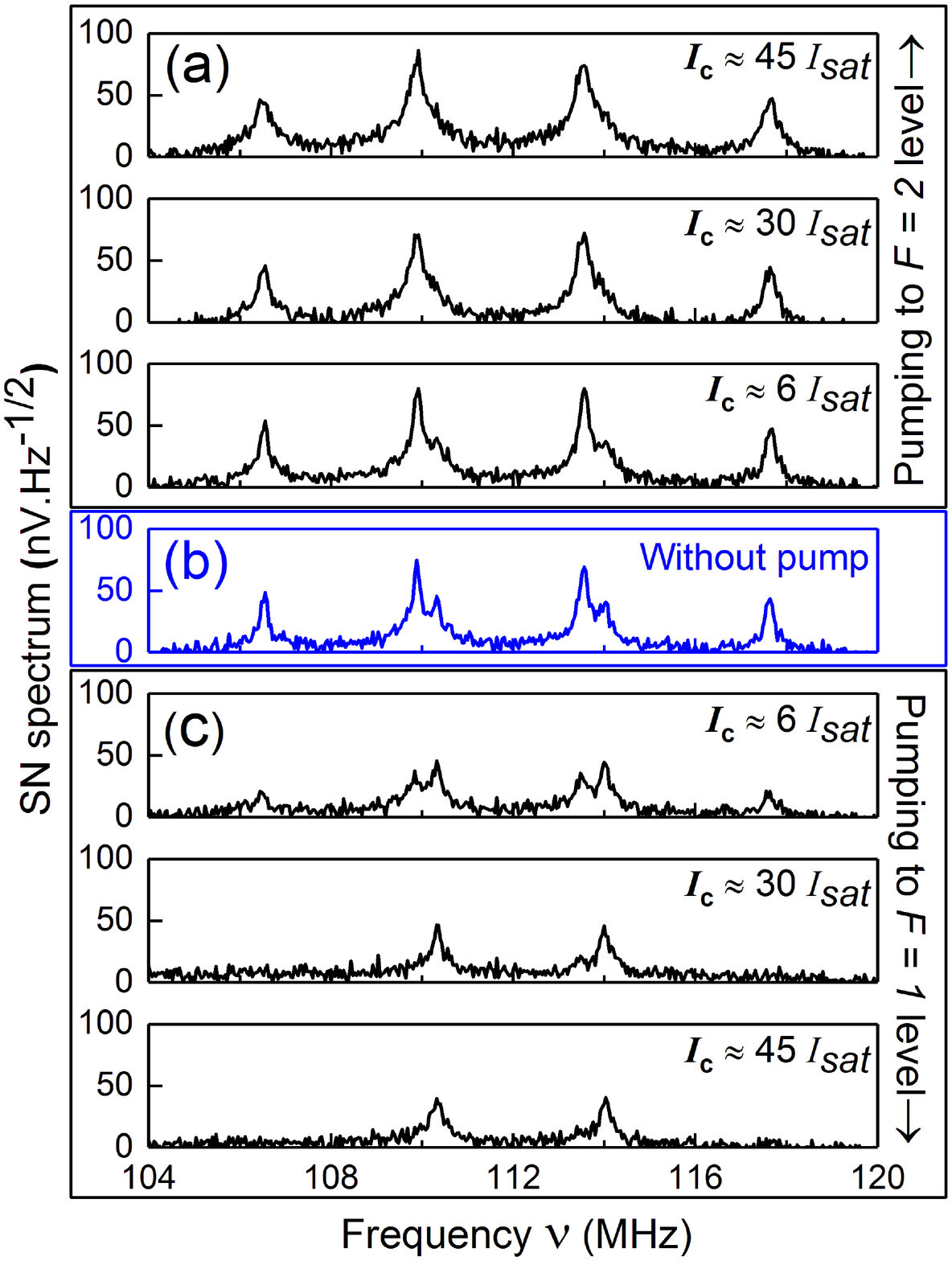}
\caption{ Spin noise (SN) spectra in and out of equilibrium $^{87}$Rb atoms. (a) SN spectra with a pump beam on resonant to $F = 1 \rightarrow F' = 2$ optical transition for various pump beam intensities $I_c$. (b) SN spectrum in thermal equilibrium (without optical pump beam). (c) SN spectra with a pump beam on resonant to $F = 2 \rightarrow F' = 2$ optical transition for different pump beam intensities $I_c$. For all panels, $B_\bot = 160$ G, $\delta_p=-10.6$ GHz and cell temperature = 105$^{\circ}$C.}
\label{fig:OpticalPumping}
\end{figure}

Here, we implement the off-resonant SNS to probe the spin imbalance in an optically driven system without applying further perturbation \cite{Happer1967}. In the absence of the pump beam, six spin coherences are seen in the SN spectrum in Fig.~\ref{fig:OpticalPumping}(b) as is expected from an ensemble of atoms with population in both the ground state hyperfine levels ($F = 1,2$). On setting the frequency $\nu_c$ of the control beam on resonance to the $F = 1 \rightarrow F' = 2$ transition of $^{87}$Rb (see Fig.~\ref{fig:Energy_Diagram}), a fraction of atoms is pumped out of the ground $F = 1$ level depending on the intensity $I_c$ of the pump. This is evident in Fig.~\ref{fig:OpticalPumping}(a) where only four SN peaks related to $F = 2, \bigtriangleup m_F = \pm 1$ are observed at the highest $I_c$. On the other hand, when the atoms are pumped out of the ground $F = 2$ level using a pump beam resonant with the $F = 2 \rightarrow F' = 2$ transition, the SN spectrum reduces to two peaks related to $F = 1, \bigtriangleup m_F = \pm 1$ spin coherences. This is shown in Fig.~\ref{fig:OpticalPumping}(c) for different pump beam intensities. This clearly shows that the spin populations in different ground state hyperfine levels is reflected in the SN spectra. Such relatively non-invasive detection of spin states in a non-equilibrium atomic system may find applications in atom interferometry \cite{Cronin2009}, atomic clocks \cite{Wynands2005} and gravimetry \cite{Bertoldi2006}.

\subsection{Resolving spectral lines}

\begin{figure}
\centering
\includegraphics[scale=0.35]{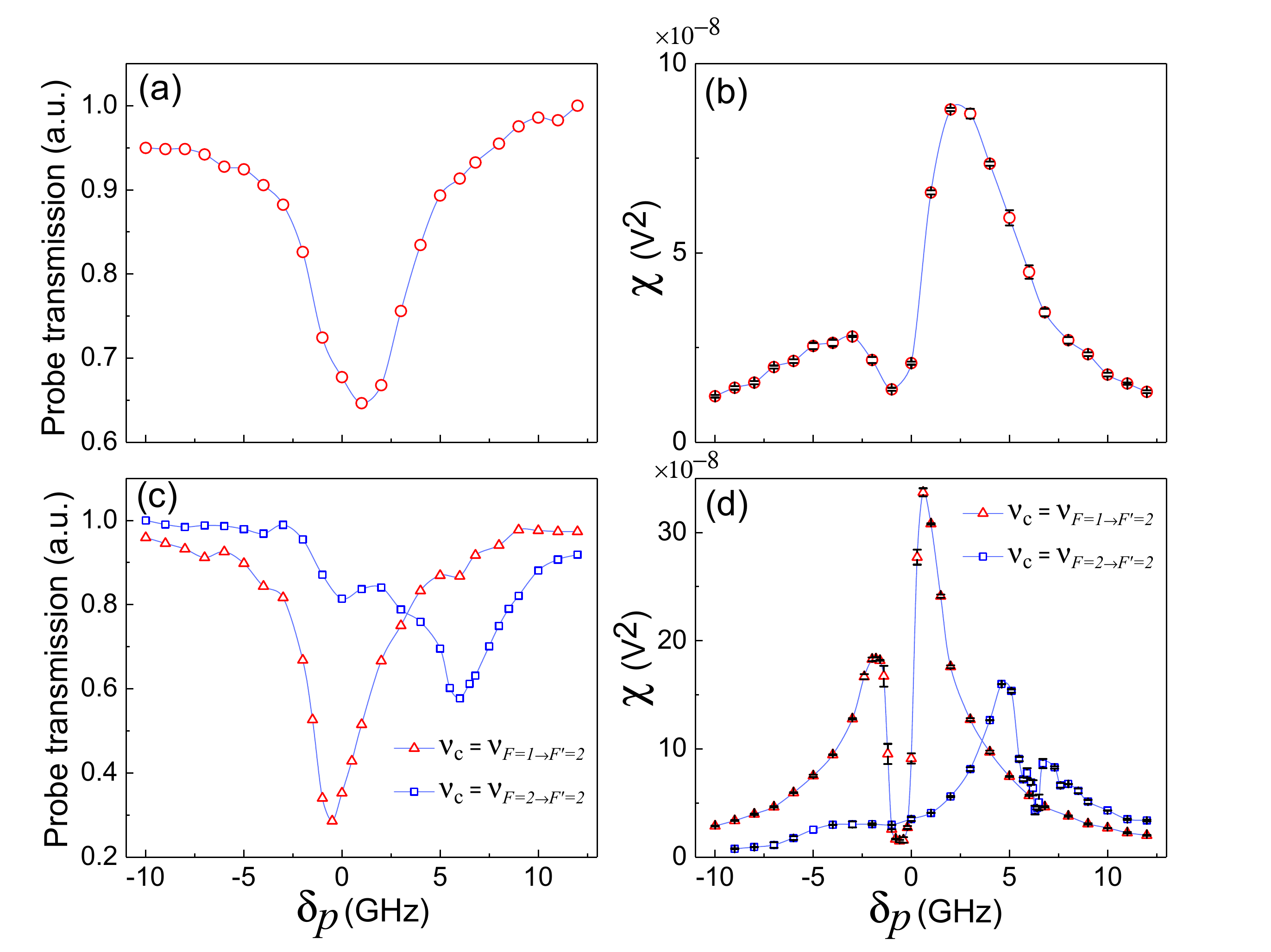}
\caption{Comparison between absorption spectroscopy and spin noise spectroscopy in resolving spectral lines in a buffer gas filled Rb vapor cell in the absence (a,b) and presence (c,d) of optical pumping. (a,c) Probe transmission  vs. probe beam detuning, $\delta_p$ defined in Fig.~\ref{fig:Energy_Diagram}. (b,d) Integrated spin noise (SN) power $\chi$ from $^{87}$Rb atoms vs. $\delta_p$. Red triangles (blue squares) depict probe transmission and integrated SN power $\chi$  from optically pumped $F=2~(F=1)$ atoms. The lines joining the data points are a guide to the eye. For all panels, $B_\bot = 7.12$ G and cell temperature = $90^{\circ}$C.}
\label{fig:SNA_detuning}
\end{figure}

The enriched $^{87}$Rb vapor cell used in our experiments contained a buffer gas (neon), and thus the conventional absorption spectra that we measure suffers broadening mainly due to homogeneous pressure broadening \cite{Romalis1997, Couture2008, Li2016} and modestly due to inhomogeneous Doppler broadening. The transmission of the probe through the atomic vapor at 90$^{\circ}$C is studied in the absence and presence of an optical control beam as shown in Fig.~\ref{fig:SNA_detuning}(a,c). The detuning $\delta_p$ of the probe beam was varied over a wide range $(-10$ GHz to $12$ GHz$)$ which covers both the $F=2 \rightarrow F'$ and $F = 1 \rightarrow F'$ transition lines. In the absence of a pump beam in Fig.~\ref{fig:SNA_detuning}(a), the probe transmission as a function of $\delta_p$ shows a single dip situated between the transition lines. The integrated SN power $\chi$ from the Rb vapor also shows a single dip in Fig.~\ref{fig:SNA_detuning}(b) in the absence of optical pumping. Thus, both the absorption spectroscopy and SNS fail to detect $F=2 \rightarrow F'$ and $F = 1 \rightarrow F'$ transition lines separately. Nevertheless, the dip in $\chi$ seems to indicate a red-detuned $F=2 \rightarrow F'$ transition as expected in the presence of neon buffer gas \cite{Li2016}.   

In the case of an optically pumped atomic ensemble, the probe transmission (Fig.~\ref{fig:SNA_detuning}(c)) can detect the above two optical transitions independently. In Fig.~\ref{fig:SNA_detuning}(d), we show the integrated SN power $\chi$ with the probe detuning $\delta_p$ when the atoms are optically pumped to either $F=2$ or $F=1$ level by a pump intensity $I_c > 50 I_{\text sat}$. We observe a dip in each integrated SN power near $F=2 \rightarrow F'$ or $F = 1 \rightarrow F'$ transition lines. Therefore, the SNS can also be used to resolve the spectral lines in a driven atomic system \cite{Zapasskii2013b}. Moreover, SNS has a better resolution (around three times in Fig.~\ref{fig:SNA_detuning}(d) than Fig.~\ref{fig:SNA_detuning}(c)) over the absorption spectroscopy \cite{Zapasskii2013b}.

\begin{figure}
\centering
\includegraphics[scale=0.38]{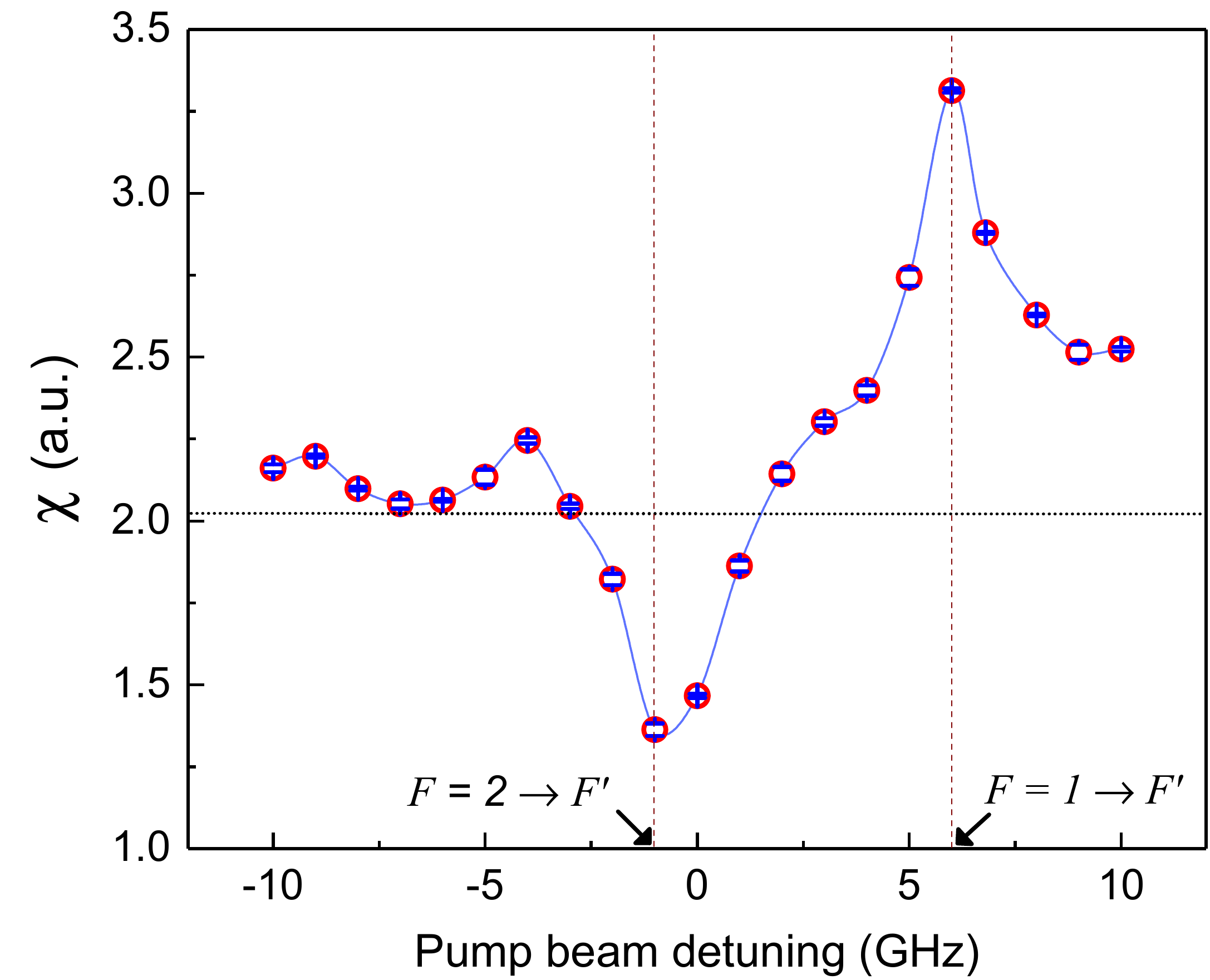}
\caption{Resolving spectral lines of $^{87}$Rb atoms in the presence of neon buffer gas by tuning the pump beam frequency $\nu_c$. Here, the probe beam detuning $\delta_p = -10$ GHz and the pump beam intensity $I_c \sim 50 I_{\text sat}$. The black dotted horizontal line represents the integrated spin noise power $\chi$ without optical pumping. The line joining the data points are a guide to the eye.
}
\label{fig:PumpBeamDiffDetuning}
\end{figure}

We can also detect these transitions by tuning the frequency $\nu_c$ of the pump beam instead of the probe beam. Here we keep the probe detuning $\delta_p$ fixed at $-10$ GHz from $F = 2 \rightarrow F'$ transition (and around $-16.8$ GHz detuned from $F = 1 \rightarrow F'$ transition). Therefore, most of the contribution in the SN signal comes from the $F = 2$ level. We tune the frequency $\nu_c$ of the pump beam from $-10$ GHz to 10 GHz around $F = 2 \rightarrow F'=3$ transition. We plot the integrated SN power $\chi$ as a function of the pump beam detuning in Fig.~\ref{fig:PumpBeamDiffDetuning}, and observe a clear dip near $F = 2 \rightarrow F'$ transition and a prominent peak near $F = 1 \rightarrow F'$ transition. Therefore, unlike conventional spectroscopy, in the case of SNS, we have the freedom to scan the pump beam for detecting the spectral lines instead of applying the pump beam at a particular known frequency as in Fig.~\ref{fig:SNA_detuning}(c,d). This, we believe, will be of particular advantage, when we wish to probe local environment-induced energy level shifts, or in resolving ground state levels in complex molecular and condensed matter systems where one has incomplete knowledge of energy levels.

\section{Conclusion and Outlook}
\label{conc}

We explore the SNS technique in atomic vapor of Rb in thermal equilibrium as well as in a system driven out of equilibrium by optical pumping. Optical pumping is a commonly used technique in atomic and optical science and technology. To the best of our knowledge, the present study is the first implementation of SNS in an optically pumped atomic system. Our effort to combine the SNS and the optical pumping has the potential for use in magnetometry with alkali atoms \cite{Allred2002}. There are important applications of such {\it{precision magnetometry}}, e.g., in cold atom experiments where narrow Feshbach resonances \cite{Chin2010} are used extensively with the resonances occurring at magnetic fields ranging from a few Gauss to a few hundred Gauss. In those experiments, measuring the external fields with ultra-high precision will be hugely beneficial in fixing the interaction energy scale. We also extend the applicability of SNS in precision measurements of various atomic, nuclear and magnetic properties in equilibrium systems.

The relatively non-perturbative nature of SNS makes it a versatile non-invasive detection technique which can be utilized in a wide range of physical systems in atomic, molecular and condensed matter systems. Recently, some measurements of spin polarization in ultracold atoms via Faraday rotation of a far-detuned probe beam were carried out as a non-destructive imaging technique \cite{Gajdacz2013, Gajdacz2016, Yamamoto2017, Palacios2018}. We are interested to further implement SNS using Faraday rotation noise in ultracold atoms and Bose-Einstein condensates where it may have significant application in quantum non-demolition measurements. However, acquiring sufficient time-resolved Faraday-rotation noise from such ultracold atomic systems poses a challenge.

\begin{acknowledgements}
The authors acknowledge funding support from Department of Science and Technology, India. The authors also acknowledge the contribution of Meena M. S. for the help with electronics and mechanical workshop of Raman Research Institute for the hardware development. D.R. acknowledges funding from the Department of Science and Technology, India, via the Ramanujan Fellowship.

\end{acknowledgements}

\bibliography{OEbibliographySNS} 
  
\end{document}